# Tunable Magnon Polaritons via Eddy-Current-Induced Dissipation in Metallic-Banded YIG Spheres

Tatsushi Uno[1], Shugo Yoshii[1] [†], Sotaro Mae[1], Ei Shigematsu[1,2,3] [‡], Ryo Ohshima[1,3], Yuichiro Ando[1,3,4], and Masashi Shiraishi[1,3*]

[1] Department of Electronic Science and Engineering, Kyoto University, Kyoto 615-8510, Japan

[2] Institute for Chemical Research, Kyoto University, Uji 611-0011, Japan

[3] Center for Spintronics Research Network, Kyoto University, Uji 611-0011, Japan

[4] PRESTO, Japan Science and Technology Agency, Saitama 332-0012, Japan

[†] yoshii.shugo.x63@kyoto-u.jp

[‡] shigematsu.ei.2r@kyoto-u.ac.jp

[*] shiraishi.masashi.4w@kyoto-u.ac.jp

Abstract

We demonstrate a robust method to dynamically tune magnon dissipation in yttrium iron garnet spheres by equipping a metallic band around the sphere's equator, enabling precise control over magnon-photon coupling states. The collective magnetization dynamics in the YIG sphere induce circular eddy currents in the metallic band, whose magnitude can be systematically varied by adjusting the angle between the metallic band plane and an external static magnetic field. This angular dependence yields a pronounced modulation of the ferromagnetic resonance (FMR) linewidth, facilitating seamless transitions between the Purcell and strong coupling regimes without altering photon cavity parameters. Systematic FMR and cavity spectroscopy measurements confirm that eddy-current-induced losses govern the primary mechanism behind the observed tunable damping. By achieving extensive $\cos^2\theta$-type tunability of magnon relaxation rate, we precisely control the magnon-photon coupling state, approaching the critical coupling condition ($g=\gamma/2$). These results establish the YIG-metallic-band platform as a versatile and practical approach for engineering tunable magnon-polariton systems and advancing magnonic applications, including those exploring non-Hermitian magnonics.

Magnons—quantized spin waves—have recently emerged as one of the central physical objects in condensed matter physics, offering a rich platform for both fundamental physics exploration and practical applications. Over the past few decades, magnons have attracted considerable attention for constructing quantum hybrid systems, owing to their robust capacity for energy exchange with other quantum information media, including photons [1–3], phonons [4,5], and additional magnon modes [6,7]. Of particular interest is the strong magnon-photon coupling that gives rise to magnon-polaritons, fueling advances in quantum memory [8–10] and quantum transduction [11]. A crucial milestone in the study of magnon polaritons has been achieving controllable transitions between weak and strong coupling regimes, accomplished through temperature-dependent magnetization [12] and precise tuning of the coupling via photon mode [13]. These developments have further enabled exploration into novel phenomena represented by non-Hermitian magnon-photon polariton systems [14–20].

Central of much of this progress is the ferrimagnetic insulator yttrium iron garnet ($Y_3Fe_5O_{12}$, YIG). Spherical YIG, characterized by ultralow damping and bulk ferromagnetic properties, have been instrumental in advancing strong magnon-photon coupling research [3,9,21–26], significantly enriching our understanding of magnon-photon interactions. However, while the large magnetic volume of millimeter-scale YIG spheres is advantageous for achieving strong coupling, it also predominantly ties the magnon mode to intrinsic magnetization states governed by the external magnetic field. This dependence complicates the incorporation of additional control mechanisms, thereby limiting the versality of these systems in tunable magnon-polariton applications. Temperature-based tuning of magnetization [12], although expected as a potential mechanism, inadvertently modifies the parameters of coupled photon modes, thereby complicating device integration and precise system coordination. Alternatively, surface-sensitive control methods—such as spin pumping, modulation of interfacial magnetic anisotropy [27–32], or spin-torque applications via the spin Hall effect [33,34]—offer finer spatial resolution but remain fundamentally incompatible with large-scale YIG resonators due to their limited effective range on the nanoscale. Thus, the lack of a broadly applicable, scalable approach for tuning magnons in large YIG spheres remains a significant obstacle hindering further advancements in magnon-polariton-based technologies.

To overcome the challenge of robust magnon control in large-scale YIG spheres, we introduced a metallic band structure encircling the equator of the sphere, as illustrated in Fig. 1. Eddy currents induced within this metallic band by collective magnetization dynamics are expected to strongly couple to the extensive magnon volume present in the YIG sphere [35,36], thus providing a promising mechanism for effective magnon manipulation. By systematically rotating the YIG sphere to adjust the orientation of the metallic band relative to the external static magnetic field, we achieved precise modulation of the dissipation rate of uniform magnons. This dissipation exhibited a distinct angular dependence, clearly arising from eddy currents generated by uniform magnons within the metallic structure rather than conventional surface-based phenomena such as spin pumping or interfacial effects. Crucially, our approach maintains photon mode parameters unchanged, enabling dynamic transitions between the Purcell regime and the strong-coupling regime in magnon-polariton systems. Our finding significantly advances the integration potential of YIG spheres within sophisticated magnonic platforms, particularly those exploring the physics of non-Hermitian magnon-polariton systems.

To investigate the effects of adjacent metallic bands on magnon damping, we prepared three types of YIG spheres: a bare YIG sphere (sample-bare), a YIG sphere with a $SiO_2$/Cu/$SiO_2$ band (sample-$SiO_2$/Cu), and a YIG sphere with a Pt band (sample-Pt). Sample-$SiO_2$/Cu was designed to assess the impact of the metallic band as a conductor on magnon modes while eliminating spintronic effects such as spin pumping or the inverse spin Hall effect. This was achieved by using Cu, a material with weak spin-orbit interaction, and preventing direct contact between the metallic band and the YIG sphere through an insulating $SiO_2$ layer. Sample-Pt, in contrast, was used in photon-magnon coupling experiments because the SiO$_2$/Cu structure is mechanically fragile, and Cu degrades under prolonged atmospheric exposure.

Commercially available single crystalline YIG spheres, having a uniform diameter of 1 mm, were used in this study. All YIG spheres were grown via liquid-phase epitaxy and had a uniform diameter of 1 mm. The metallic bands were deposited on the YIG spheres via sputtering under a base pressure of less than 5 × 10$^{-5}$ Pa, with an Ar gas pressure of 0.5 Pa during plasma application. The bands were patterned using a lift-off process with PMMA resist to ensure precise formation.

The influence of the metallic bands on magnons excited within the YIG spheres was studied by measuring the absorption spectra of the samples placed on a coplanar waveguide (CPW), as shown in Fig. 2(a). The YIG spheres were mounted on the center conductor of the CPW, with the metallic bands oriented parallel to the waveguide plane. A static magnetic field $H$ was applied either in-plane (IP) or out-of-plane (OOP) with respect to the CPW. A 0 dBm microwave signal, sweeping frequencies from 0.02 GHz to 9 GHz in 0.02 GHz increments, was supplied via a vector network analyzer (VNA) to generate an AC magnetic field on the CPW. The transmission parameter $S_{21}$ was recorded while sweeping the static magnetic field.

The relationship between the susceptibility ($\chi$) measured in the CPW-VNA setup and the normalized deviation ($\Delta S_{21}$) is given by $\Delta S_{21} = -\mathrm{j}A\omega\chi$, where $A$ and $\omega$ are the amplitude constant and angular frequency of the AC magnetic field, respectively. For a ferromagnetic sphere with a metallic band, the susceptibility ($\chi$) can be expressed as:

$$\chi = \frac{\mu_0 M(\mu_0 H + \mathrm{j}\mu_0 \Delta H)}{(\mu_0 H + \mathrm{j}\mu_0 \Delta H)^2 - \omega^2/\gamma^2}, \qquad (1)$$

where $\mu_0$ is the vacuum permeability, $\mu_0 M$ the saturation magnetization, $\mu_0 H$ the static magnetic field, $\mu_0 \Delta H$ the half-width at half-maximum (HWHM) of the FMR spectra, and $\gamma$ the gyromagnetic ratio. Figure 2(b)-(g) shows the extracted FMR spectra from the CPW-VNA measurements at a 2.6 GHz microwave frequency.

While the FMR spectra of all samples exhibited similar shapes in the out-of-plane (OOP) configuration, samples with metallic bands (SiO$_2$/Cu/SiO$_2$ and Pt) displayed significantly broader linewidths in the in-plane (IP) configuration compared to the OOP configuration. The $\mu_0\Delta H$ values obtained by fitting Eq. (1) for sample-bare, sample-SiO$_2$/Cu, and sample-Pt in the OOP configuration are 0.97 ± 0.32 mT, 1.42 ± 0.34 mT, and 0.94 ± 0.42 mT, respectively. However, in the IP configuration, $\mu_0\Delta H$ is enhanced to 1.78 ± 0.14 mT for sample-SiO$_2$/Cu and 1.74 ± 0.13 mT for sample-Pt, whereas no enhancement is observed for sample-bare. This comparable enhancement in $\mu_0\Delta H$ for the two metallic band samples suggests that interfacial damping effects, such as spin pumping due to the Pt band, are not the primary contributors.

Instead, the observed damping enhancement in the IP configuration is likely due to eddy currents induced in the metallic bands by the electromagnetic induction effect arising from the magnetization dynamics of the YIG sphere. Considering the spin pumping-induced current and eddy currents caused by YIG magnetization dynamics, the susceptibility in the IP configuration is expressed as:

$$\chi = \frac{\mu_0 M \left[\mu_0 H + \frac{j\omega\alpha}{\gamma} + j\omega\beta\mu_0 M - \omega\varepsilon\mu_0 M\right]}{\left[\mu_0 H + \frac{j\omega\alpha}{\gamma}\right]\left[\mu_0 H + \frac{j\omega\alpha}{\gamma} + j\omega\beta\mu_0 M - \omega\varepsilon\mu_0 M\right] - \frac{\omega^2}{\gamma^2}}, \quad (2)$$

where $\alpha$, $\beta$, and $\varepsilon$ represent the Gilbert damping parameter in YIG sphere, the contributions of eddy current effects and the possible inverse spin Hall current induced by spin pumping, respectively. Notably, the eddy current contributes to linewidth broadening, while the spin pumping-induced current causes a resonance frequency shift in the FMR spectra (detailed in Supplemental Material [37]). These findings strongly indicate that the modulation of magnon relaxation in metallic band-adjacent YIG spheres primarily originates from the circular eddy currents excited by the magnetization dynamics.

The magnetic angular dependence of $\mu_0\Delta H$ was experimentally determined by measuring the FMR linewidth as the angle of the external magnetic field was rotated in a plane perpendicular to the metallic band, as illustrated in Fig. 3(a). In this configuration, the uniform $TE_{011}$ microwave photon mode strongly excites the Kittel mode of the YIG sphere. The samples were placed in a rectangular resonator, with the static magnetic field swept along the $-y$ direction. The microwave frequency and power were fixed at 9.12 GHz and 0 dBm, respectively. The rotation angle ($\theta$) for samples with metallic bands was defined as the angle between the metallic band and the static magnetic field.

Figures 3(b) and 3(c) depict the FMR spectra for sample-bare and sample-$SiO_2$/Cu at $\theta = 0°$ and $90°$. While the FMR spectra of sample-bare remain largely unchanged, the linewidth of the spectra for sample-$SiO_2$/Cu is significantly modulated as $\theta$ changes. The spectra can be described using the following equation [31]:

$$I = \frac{S(\mu_0 \Delta H)^2}{(\mu_0 H - \mu_0 H_{res})^2 + (\mu_0 \Delta H)^2} + \frac{A(\mu_0 H - \mu_0 H_{res})\mu_0 \Delta H}{(\mu_0 H - \mu_0 H_{res})^2 + (\mu_0 \Delta H)^2}, \qquad (3)$$

where $I$, $\mu_0 H$, $\mu_0 H_{res}$, $\mu_0 \Delta H$, $S$, and $A$ are intensity of spectra, the static magnetic field, the resonance field, the HWHM of FMR, symmetric and asymmetric components, respectively. For sample-bare, the estimated $\mu_0 \Delta H$ values are $1.05 \pm 0.00$ mT ($\theta = 0°$) and $1.32 \pm 0.00$ mT ($\theta = 90°$). In contrast, sample-$SiO_2$/Cu shows significantly different values: $7.07 \pm 0.04$ mT ($\theta = 0°$) and $0.91 \pm 0.04$ mT ($\theta = 90°$). The enhanced damping for the $k = 0$ magnon mode, where the precession angle is parallel to the band plane, aligns with the experimental results obtained using the CPW. This finding highlights the key role of angular dependence in determining the dissipation properties of magnons in the presence of metallic bands.

Figure 3(d) presents the angular dependence of the estimated $\mu_0 \Delta H$ for both the bare sample and the $SiO_2$/Cu sample using Eq. (3) as the fitting function of the FMR spectra. Notably, the magnon damping in the YIG sphere with Cu/$SiO_2$ is strongly modulated by varying the angle of the external magnetic field, whereas the FMR linewidth of the bare sample remains constant. Based on the CPW-FMR results, this pronounced modulation can be attributed to eddy-current loss in the metallic band. To clarify the mechanism underlying the modulation of magnon dissipation modulation, we examine the FMR linewidth as a function of the magnetic field orientation from both experimental and theoretical perspectives. By considering the circular current induced in the metallic band due to the tilting of the magnetization dynamics toward the band plane, the half-width at half maximum (HWHM) $\mu_0 \Delta H$ is expressed as (see Supplemental Material [36] for details of the derivation):

$$\mu_0 \Delta H = \frac{\omega \alpha}{\gamma} + \frac{\omega \beta \mu_0 M_z \cos^2 \theta}{2}. \qquad (4)$$

In this framework, the FMR linewidth—including the eddy-current effect generated by magnons in the YIG sphere—is described by Eq. (4) with explicit $\theta$-dependence. The extracted parameter $\beta$ for the SiO$_2$/Cu sample is $1.22 \times 10^{-12}$ s/rad, which is comparable to the value obtained from the CPW-FMR measurement ($0.80 \times 10^{-12}$ s/rad) through fitting Fig. 2(g) with Eq. (2).

Invoking Ampère's law for the inductive circular current excited in the metallic band by uniform magnetization dynamics, one obtains the relation $\beta = \frac{\pi r \mu_0}{2R}$, where $r$ and $R$ denote the radius and resistance of the metallic band, respectively. A typical two-terminal $I$-$V$ measurement of the metallic band yields $R$ as 356 Ω (see Supplemental Materials [37]), leading to $\beta = 2.77 \times 10^{-12}$ s/rad. The close agreement of $\beta$ values from independent measurements provides strong evidence that the circular ac current in the metallic band effectively suppresses magnons in the adjacent YIG sphere. Furthermore, rotating the magnetic field relative to the plane of the metallic band modulates the interaction between the circular current and the magnetization dynamics in the YIG sphere, thus offering a promising route for the fine-tuning of magnon damping.

The magnon–photon coupling state is characterized by the coupling strength $g$, the magnon relaxation rate $\gamma$, and the photon relaxation rate $\kappa$ [2]. Because $\gamma$ in a YIG sphere with a metallic band can be controlled by changing the angle of the external magnetic field, as demonstrated in the experiments above, such a YIG sphere serves as a key component for achieving tunable strong magnon–photon coupling. In particular, the circular eddy current in the metallic band adjacent to the YIG sphere—rather than spintronic spin-to-charge conversion—can significantly affect the magnon dynamics. Therefore, we employed a Pt-containing sample ("sample-Pt") to demonstrate finely tunable magnon–photon coupling.

A TiO$_2$ (rutile) cylindrical resonator [38] was used to mitigate photon losses originating from the cavity. As shown in Fig. 4(a), the sample–Pt assembly, attached to a rotatable rod, was inserted into the rutile resonator with a designed resonance frequency of 2.61 GHz. The resonator was mounted on a sapphire substrate and placed in a Cu shield to suppress radiation losses. The microwave reflection $S_{11}$ was measured via a circular coil antenna adjacent to the rutile resonator, while a static magnetic field was applied by an electromagnet. The angle $\theta$ between the external magnetic field and the plane of the metallic band was controlled by rotating the rod supporting the YIG sphere.

Figures 4(b)–4(d) present the magnitude of $|S_{11}|$ as a function of frequency and applied magnetic field at $\theta = 0°$, 45°, and 90°, respectively. In each spectrum, a splitting appears at a certain value of the magnet current, indicating an interaction between the magnon mode in the YIG sphere and the photon mode in the rutile resonator. The variation in the resonant current values stems from the shift in the magnetic resonance field due to the crystalline magnetic anisotropy of YIG (see Supplemental Material [37]). Notably, the pronounced anti-crossing observed at $\theta = 90°$ contrasts with the nearly absent anti-crossing at $\theta = 0°$. This behavior is consistent with the experimental observation that magnon damping is significantly enhanced when the external magnetic field is applied parallel to the band plane.

Figure 4(e) shows the $\theta$-dependence of $|S_{11}|$ around the magnon resonance current. While only a single peak appears up to $\theta = 45°$, the peak splits into two from $\theta = 45°$ to 90°. This transition signifies the onset of strong magnon–photon coupling as the external magnetic field orientation is tilted relative to the plane of the band.

To determine the coupling strength and relaxation rates of the magnon and photon modes, we employ the standard input-output theory [39]. In this framework, the microwave reflection

coefficient is given by

$$S_{11} = \frac{i(\omega - \omega_c) - \frac{\kappa_i - \kappa_e}{2} + \frac{g^2}{i(\omega - \omega_m) - \frac{\gamma}{2}}}{i(\omega - \omega_c) - \frac{\kappa_i + \kappa_e}{2} + \frac{g^2}{i(\omega - \omega_m) - \frac{\gamma}{2}}}, \qquad (5)$$

where $\omega_c$ and $\omega_m$ are the resonance angular frequencies of the cavity and ferromagnetic resonance (FMR) mode, respectively, while $\kappa_i$ and $\kappa_e$ represent the cavity's internal dissipation and external coupling rates. By fitting the experimental spectra at $\theta = 0°$, $45°$ and $90°$ to Eq. (5), we obtain the system parameters summarized in Table 1.

Based on these fitted parameters, the magnon–photon coupling states at different angles $\theta$ can be classified as follows: the Purcell regime ($\theta = 0°$), near the compensating point ($g \sim \gamma/2$) for $\theta = 45°$, and the strong-coupling regime ($\theta = 90°$). These assignments follow the criteria outlined in Ref. [2] for distinguishing between different coupling states. Notably, the ability to realize these regimes simply by rotating the YIG sphere, which is surrounded by a metallic band, highlights the robust tunability of this system. Although the spectra in Fig. 4(e) exhibit a single peak from $\theta = 0°$ up to approximately $\theta = 45°$, a progressive evolution into a split-peak structure emerges near $\theta = 90°$. This gradual transition underscores the wide tuning range of the coupling conditions in this system. As discussed, this tunability stems from the angular dependence of the magnon relaxation rate $\gamma$, which is expected to follow a $\cos^2\theta$ -type behavior.

Figure 5 shows the $\theta$-dependence of $g$ and $\gamma/2$, extracted from fittings to Eq. (5). Because the total cavity dissipation rate $\kappa/2\pi = (\kappa_i + \kappa_e)/2\pi$ remains below 2 MHz at all angles (see Supplemental Material [37]), the coupling regime is classified as either the Purcell-effect regime ($\gamma/2 > g > \kappa/2$) or the strong-coupling regime ($g > \kappa/2, \gamma/2$). Interestingly, while $g$ remains nearly constant with respect to $\theta$, $\gamma/2$ exhibits a clear $\cos^2\theta$ dependence, allowing the system to be tuned precisely to the condition $g = \gamma/2$. This fine-tuning capacity is promising for constructing platforms dedicated to non-Hermitian physics.

Furthermore, $\gamma$ can be related to the half-width at half-maximum (HWHM) in FMR through the dispersion relation $\omega/2\pi = \gamma\mu_0|H|$. We therefore apply the fitting procedure described in Eq. (4) to the $\gamma$ data in Fig. 5, yielding an excellent agreement and an extracted $\beta = 1.63 \pm 0.03 \times 10^{-12}$ s/rad, consistent with results from ESR-FMR measurements. These findings corroborate highly effective modification of the magnon dissipation in a YIG sphere by simply adjusting the orientation of the external static magnetic field relative to the metallic band. This technique achieves sufficient tunability to traverse multiple coupling regimes, thereby providing a valuable tool for realizing long-anticipated milestones in non-Hermitian magnonics.

Before concluding, we turn our attention to possible future developments for this structure. The demonstrated method of controlling magnon dissipation in a YIG sphere can also be viewed as modulating the coupling strength between the metallic band and the YIG sphere. Consequently, this opens the door to introducing tunable coupling strength with switchable transducer functionality in the ring resonator–YIG sphere system proposed in Refs. [23,40]. Moreover, since $\beta$ is inversely proportional to the metallic band's resistance, exploiting the ultralow resistance of a superconductor provides a route to increase $\beta$ significantly. This enhancement broadens the modulation range and thus enables wide tunability of the magnon–polariton coupling state in the YIG sphere.

In conclusion, we have shown that by rotating the relative angle of the external magnetic field to a metallic band on a YIG sphere, the magnon–photon coupling state can be tuned between the Purcell and strong coupling regime. This approach modulates the magnon relaxation rate $\gamma$ via eddy currents in the metallic band, yielding a $\cos^2\theta$-type sinusoidal control of magnon damping while maintaining a constant coupling strength. Such tunability, combined with the inherently low damping of YIG, establishes a versatile platform for exploring non-Hermitian magnon-polariton physics, where the engineered manipulation of magnon dissipation is essential for advanced magnonic functionalities.

Acknowledgements

The authors thank the financial support from Japan Society for the Promotion of Science (KAKENHI Grant No. 22K14301), The Murata Science Foundation, JSPS Research Fellow Program (grant no. 22KJ1956) and the Spintronics Research Network of Japan. TU, SY, and ES thank Haruka Komiyama and Ryusuke Hisatomi for the fruitful discussions about this work.

[1] H. Huebl, C. W. Zollitsch, J. Lotze, F. Hocke, M. Greifenstein, A. Marx, R. Gross, and S. T. B. Goennenwein, High Cooperativity in Coupled Microwave Resonator Ferrimagnetic Insulator Hybrids, Phys. Rev. Lett. **111**, 127003 (2013).

[2] X. Zhang, C.-L. L. Zou, L. Jiang, and H. X. Tang, Strongly Coupled Magnons and Cavity Microwave Photons, Phys. Rev. Lett. **113**, 156401 (2014).

[3] Y. Tabuchi, S. Ishino, T. Ishikawa, R. Yamazaki, K. Usami, and Y. Nakamura, Hybridizing Ferromagnetic Magnons and Microwave Photons in the Quantum Limit, Phys. Rev. Lett. **113**, 083603 (2014).

[4] Y. Hwang et al., Strongly Coupled Spin Waves and Surface Acoustic Waves at Room Temperature, Phys. Rev. Lett. **132**, 056704 (2024).

[5] M. Müller et al., Chiral phonons and phononic birefringence in ferromagnetic metal–bulk acoustic resonator hybrids, Phys. Rev. B **109**, 024430 (2024).

[6] Y. Shiota, T. Taniguchi, M. Ishibashi, T. Moriyama, and T. Ono, Tunable Magnon-Magnon Coupling Mediated by Dynamic Dipolar Interaction in Synthetic Antiferromagnets, Phys. Rev. Lett. **125**, 017203 (2020).

[7] T. Dion et al., Ultrastrong magnon-magnon coupling and chiral spin-texture control in a dipolar 3D multilayered artificial spin-vortex ice, Nat. Commun. **15**, 4077 (2024).

[8] X. Zhang, C.-L. Zou, N. Zhu, F. Marquardt, L. Jiang, and H. X. Tang, Magnon dark modes and gradient memory, Nat. Commun. **6**, 8914 (2015).

[9] Y. Tabuchi, S. Ishino, A. Noguchi, T. Ishikawa, R. Yamazaki, K. Usami, and Y. Nakamura, Coherent coupling between a ferromagnetic magnon and a superconducting qubit, Science **349**, 405 (2015).

[10] D. Lachance-Quirion, S. P. Wolski, Y. Tabuchi, S. Kono, K. Usami, and Y. Nakamura, Entanglement-based single-shot detection of a single magnon with a superconducting qubit, Science **367**, 425 (2020).

[11] A. A. Clerk, K. W. Lehnert, P. Bertet, J. R. Petta, and Y. Nakamura, Hybrid quantum systems with circuit quantum electrodynamics, Nat. Phys. **16**, 257 (2020).

[12] H. Maier-Flaig, M. Harder, S. Klingler, Z. Qiu, E. Saitoh, M. Weiler, S. Geprägs, R. Gross, S. T. B. Goennenwein, and H. Huebl, Tunable magnon-photon coupling in a compensating ferrimagnet—from weak to strong coupling, Appl. Phys. Lett. **110**, 132401 (2017).

[13] D. Zhang, X.-Q. Luo, Y.-P. Wang, T.-F. Li, and J. Q. You, Observation of the exceptional point in cavity magnon-polaritons, Nat. Commun. **8**, 1368 (2017).

[14] T. Yu, J. Zou, B. Zeng, J. W. Rao, and K. Xia, *Non-Hermitian Topological*

*Magnonics*, Physics Reports 1062, 1-86 (2024).

[15] R. El-Ganainy, K. G. Makris, M. Khajavikhan, Z. H. Musslimani, S. Rotter, and D. N. Christodoulides, Non-Hermitian physics and PT symmetry, Nat. Phys. **14**, 11 (2018).

[16] X. Zhang, K. Ding, X. Zhou, J. Xu, and D. Jin, Experimental Observation of an Exceptional Surface in Synthetic Dimensions with Magnon Polaritons, Phys. Rev. Lett. **123**, 237202 (2019).

[17] J. Qian, J. Li, S. Y. Zhu, J. Q. You, and Y. P. Wang, Probing PT-Symmetry Breaking of Non-Hermitian Topological Photonic States via Strong Photon-Magnon Coupling, Phys. Rev. Lett. **132**, 156901 (2024).

[18] M. Harder, L. Bai, P. Hyde, and C. M. Hu, Topological properties of a coupled spin-photon system induced by damping, Phys. Rev. B **95**, (2017).

[19] G. Q. Zhang, Z. Chen, D. Xu, N. Shammah, M. Liao, T. F. Li, L. Tong, S. Y. Zhu, F. Nori, and J. Q. You, Exceptional Point and Cross-Relaxation Effect in a Hybrid Quantum System, PRX Quantum **2**, 1 (2021).

[20] C. Wang, J. Rao, Z. Chen, K. Zhao, L. Sun, B. Yao, T. Yu, Y. P. Wang, and W. Lu, Enhancement of magnonic frequency combs by exceptional points, Nat. Phys. **20**, 1139 (2024).

[21] J. Bourhill, N. Kostylev, M. Goryachev, D. L. Creedon, and M. E. Tobar, Ultrahigh cooperativity interactions between magnons and resonant photons in a YIG sphere, Phys. Rev. B **93**, 144420 (2016).

[22] J. Xu, C. Zhong, X. Han, D. Jin, L. Jiang, and X. Zhang, Coherent Gate Operations in Hybrid Magnonics, Phys. Rev. Lett. **126**, 207202 (2021).

[23] Y. Li et al., Coherent Coupling of Two Remote Magnonic Resonators Mediated by Superconducting Circuits, Phys. Rev. Lett. **128**, 047701 (2022).

[24] M. Goryachev, W. G. Farr, D. L. Creedon, Y. Fan, M. Kostylev, and M. E. Tobar, High-Cooperativity Cavity QED with Magnons at Microwave Frequencies, Phys. Rev. Appl. **2**, 054002 (2014).

[25] R. Hisatomi, A. Osada, Y. Tabuchi, T. Ishikawa, A. Noguchi, R. Yamazaki, K. Usami, and Y. Nakamura, Bidirectional conversion between microwave and light via ferromagnetic magnons, Phys. Rev. B **93**, 174427 (2016).

[26] A. A. Serga, A. V. Chumak, and B. Hillebrands, YIG magnonics, J Phys. D Appl. Phys. **43**, 264002 (2010).

[27] B. Rana and Y. Otani, Towards magnonic devices based on voltage-controlled magnetic anisotropy, Commun. Phys. **2**, 90 (2019).

[28] B. Rana, C. A. Akosa, K. Miura, H. Takahashi, G. Tatara, and Y. Otani, Nonlinear

Control of Damping Constant by Electric Field in Ultrathin Ferromagnetic Films, Phys. Rev. Appl. **14**, 014037 (2020).

[29] X.-X. Zhang, L. Li, D. Weber, J. Goldberger, K. F. Mak, and J. Shan, Gate-tunable spin waves in antiferromagnetic atomic bilayers, Nat. Mater. **19**, 838 (2020).

[30] P. Omelchenko, E. A. Montoya, C. Coutts, B. Heinrich, and E. Girt, Tunable magnetization and damping of sputter-deposited, exchange coupled Py|Fe bilayers, Sci. Rep. **7**, 4861 (2017).

[31] S. Yoshii, K. Kato, E. Shigematsu, R. Ohshima, Y. Ando, K. Usami, and M. Shiraishi, Significant suppression of two-magnon scattering in ultrathin Co by controlling the surface magnetic anisotropy at the Co/nonmagnet interfaces, Phys. Rev. B **106**, 174414 (2022).

[32] S. Yoshii, M. Müller, H. Inoue, R. Ohshima, M. Althammer, Y. Ando, H. Huebl, and M. Shiraishi, Significant modulation of Gilbert damping in ultrathin ferromagnetic films by altering the surface magnetic anisotropy, Phys. Rev. B **109**, L020406 (2024).

[33] L. J. Cornelissen, J. Liu, B. J. van Wees, and R. A. Duine, Spin-Current-Controlled Modulation of the Magnon Spin Conductance in a Three-Terminal Magnon Transistor, Phys. Rev. Lett. **120**, 097702 (2018).

[34] T. Wimmer, M. Althammer, L. Liensberger, N. Vlietstra, S. Geprägs, M. Weiler, R. Gross, and H. Huebl, Spin Transport in a Magnetic Insulator with Zero Effective Damping, Phys. Rev. Lett. **123**, 257201 (2019).

[35] R. Ohshima, S. Klingler, S. Dushenko, Y. Ando, M. Weiler, H. Huebl, T. Shinjo, S. T. B. Goennenwein, and M. Shiraishi, Spin injection into silicon detected by broadband ferromagnetic resonance spectroscopy, Appl. Phys. Lett. **110**, 182402 (2017).

[36] S. Mae, R. Ohshima, E. Shigematsu, Y. Ando, T. Shinjo, and M. Shiraishi, Influence of adjacent metal films on magnon propagation in $Y_3Fe_5O_{12}$, Phys. Rev. B **105**, 104415 (2022).

[37] UnoPRL2024_supple, (n.d.).

[38] K. Kato, R. Sasaki, K. Matsuura, K. Usami, and Y. Nakamura, High-cooperativity cavity magnon-polariton using a high-$Q$ dielectric resonator, J Appl. Phys. **134**, (2023).

[39] C. W. Gardiner and M. J. Collett, Input and output in damped quantum systems: Quantum stochastic differential equations and the master equation, Phys. Rev. A **31**, 3761 (1985).

[40] M. Song et al., Programmable Real-Time Magnon Interference in Two Remotely

Coupled Magnonic Resonators, arXiv 2309.04289 (2023).

**Table**

| $\theta$ | $g/2\pi$ (MHz) | $\gamma/4\pi$ (MHz) | $\kappa_i/4\pi$ (MHz) | Coupling state |
|---|---|---|---|---|
| 0° | 32.36 ± 0.55 | 67.05 ± 2.33 | 0.42 ± 0.01 | Purcell ($\gamma/2 > g > \kappa_i/2$) |
| 45° | 34.58 ± 0.21 | 38.43 ± 0.51 | 0.32 ± 0.01 | Compensating ($\gamma/2 \sim g > \kappa_i/2$) |
| 90° | 33.16 ± 0.03 | 5.80 ± 0.04 | 0.27 ± 0.01 | Strong ($g > \gamma/2, \kappa_i/2$) |

Table. 1 Uno *et al.*,

**Captions**

TABLE 1. Estimated system parameters: $g/2\pi$, $\gamma/4\pi$, $\kappa_i/4\pi$ obtained by fitting the data of $|S_{11}|$ at $\theta$ = 0°, 45°, 90°.

FIG. 1. Image of YIG sphere with a metallic band.

FIG. 2. (a) Schematic measurement setup for CPW-FMR. The CPW was connected to ports of the VNA. Samples were placed on the signal line of the CPW. The inset image shows the relation between the metallic band plane and magnetization direction. (b)-(g) Real parts (red circles) and imaginary parts (blue circles) of $\Delta S_{21}$ at 2.6 GHz in (b)-(d) OOP magnetic field configuration and (e)-(g) IP magnetic field configuration, where solid lines are fitting lines expressed by Eq. (1) and the estimated $\mu_0 \Delta H$ are shown. The measured samples are (b)(e) sample-bare, (c)(f) sample-SiO$_2$/Cu, and (d)(g) sample-Pt, respectively.

FIG. 3. (a) Schematic image of FMR measurement using ESR cavity (ESR-FMR), where $\theta$ is the angle between the external magnetic field and the metallic band plane. (b), (c) FMR spectra of $\theta$ = 0° and 90° for sample-bare and -SiO$_2$/Cu, respectively. (d) $\theta$-dependence of the estimated $\mu_0 \Delta H$ for sample-bare and -SiO$_2$/Cu.

FIG. 4. (a) Experimental setup for manipulation of magnon-polariton in YIG sphere. The rutile cavity was placed in a Cu shield to prevent the radiation loss of microwave. Sample-Pt attached with the ceramic rod was inserted into the rutile resonator, and spectra were measured with VNA while rotating sample-Pt by the rod The rotation angle $\theta$ is shown in the inset. (b)-(d) Two dimensional maps of $|S_{11}|$ as a function of coil current and frequency. (b), a clear anti-crossing is observed (strong coupling) at $\theta$ = 90° (d). (c) shows the process of that transition. (e) $|S_{11}|$ at the coil current value of the resonant

condition for each angle. The red data and black lines represent the fitting range and lines, respectively.

FIG. 5. $\theta$-dependence of $g/2\pi$ and $\gamma/4\pi$ of magnon-polaritons in YIG sphere. The coupling state is considered to be in Purcell effect in the range approximately from -50° to 50°, and strong coupling in the other range. The black line shows the fitting curve with the translated Eq. (4) in frequency range.

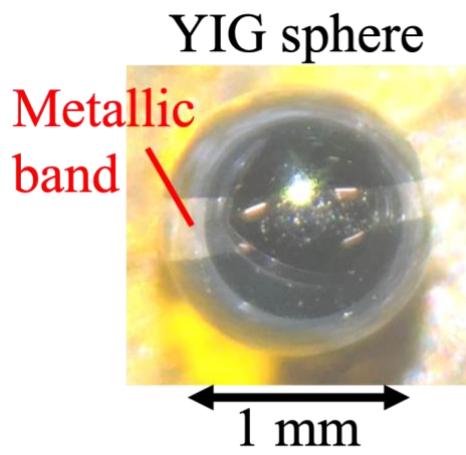

Fig. 1 Uno *et al.*,

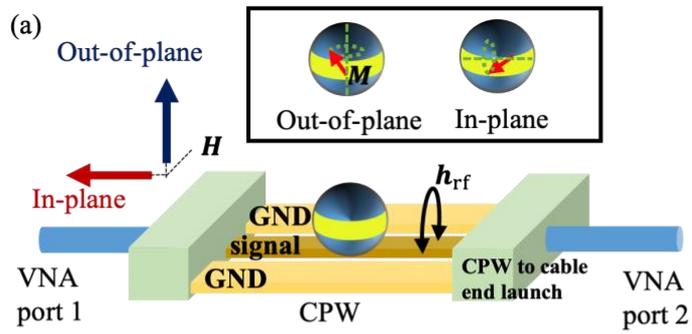
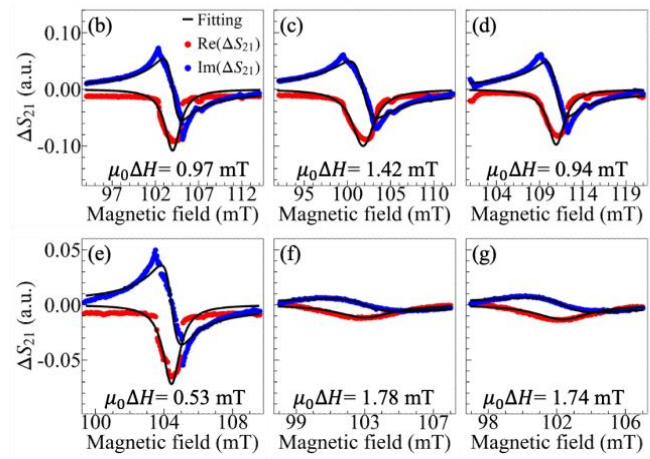

Fig. 2 Uno *et al.*,

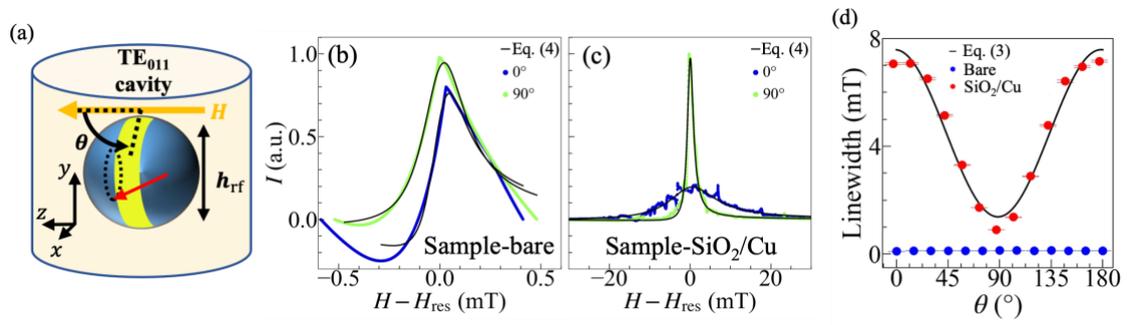

Fig. 3 Uno *et al*.,

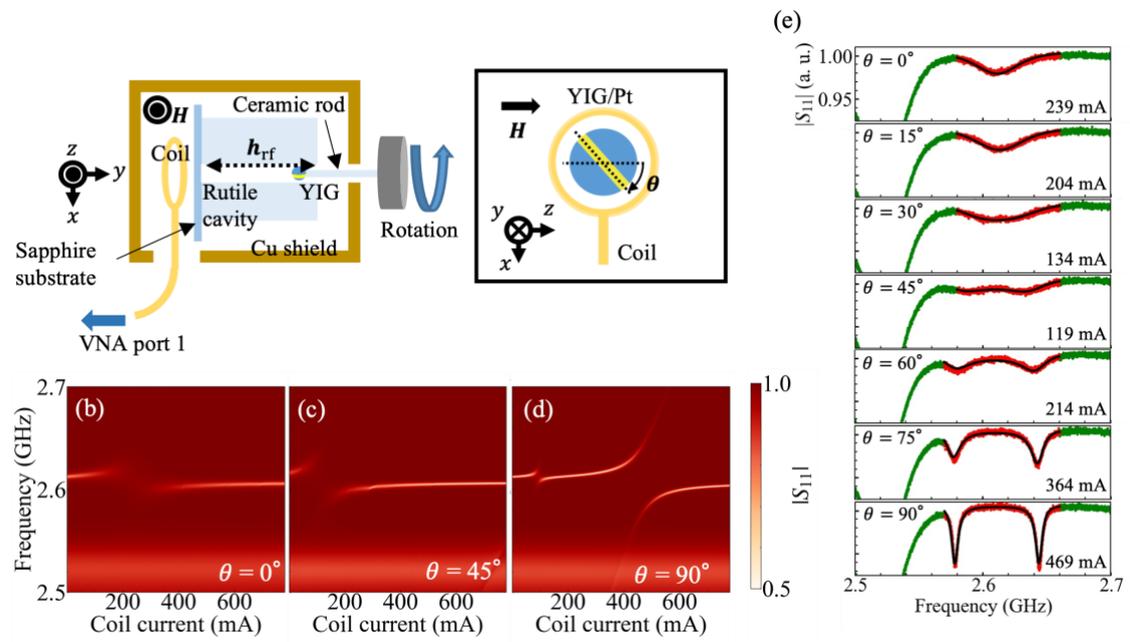

Fig. 4 Uno *et al.*,

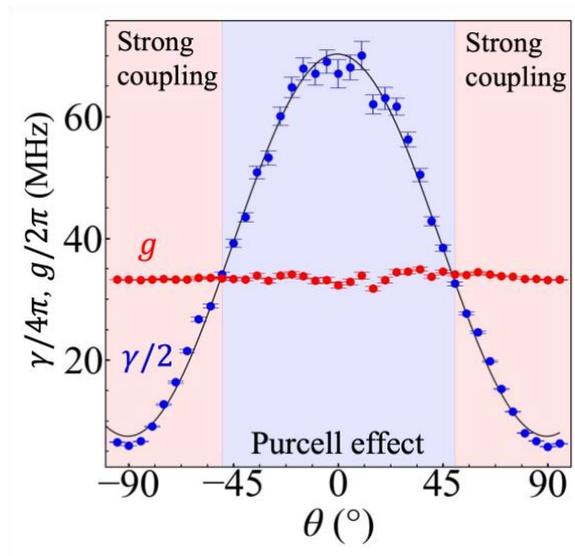

Fig. 5 Uno *et al*.